\documentclass{aa}

\usepackage{times}
\usepackage{graphics}
\begin{document}

\title{{\it Letter to the Editor}\\
ORFEUS II echelle spectra: 
detection of H$_2$\ absorption in SMC gas}

\author{
P.\,Richter\inst{1} \and
H.\,Widmann\inst{2} \and
K.S.\,de\,Boer\inst{1} \and
I.\,Appenzeller\inst{3} \and
J.\,Barnstedt\inst{2} \and
M.\,G\"{o}lz\inst{2} \and
M.\,Grewing\inst{4} \and
W.\,Gringel\inst{2} \and
N.\,Kappelmann\inst{2} \and
G.\,Kr\"{a}mer\inst{2} \and
H.\,Mandel\inst{3} \and
K.\,Werner\inst{2}
}

\institute{
Sternwarte, Universit\"at Bonn, Auf dem H\"ugel 71, D-53121 Bonn, Germany
\and
Institut f\"ur Astronomie und Astrophysik, Abt. Astronomie, Universit\"at T\"ubingen,
Waldh\"auserstr. 64, D-72076 T\"ubingen, Germany 
\and
Landessternwarte Heidelberg, K\"{o}nigstuhl, D-69117 Heidelberg, Germany
\and
Institut de Radio Astronomie Millim\'{e}trique (IRAM), 300 Rue de la Piscine, F-
38406 Saint
Martin d'H\`{e}res, France
}

\date{Received 1 July 1998 / Accepted 10 Aug. 1998 }

\thesaurus{03.19.2, 08.09.2, 11.09.4, 11.13.1, 13.21.3}

\offprints{prichter@astro.uni-bonn.de}

\titlerunning{Detection of H$_2$\ in SMC gas}

\maketitle

\begin{abstract}

We present a study of H$_2$\ in the SMC gas, based on Far UV 
spectroscopy in the line of sight to the SMC star HD\,5980.
17 absorption lines from the Lyman band have been
analysed. 
Our line of sight crosses two clouds within the SMC.
We detect a cool molecular component near +120 km\,s$^{-1}$, where
the H$_2$\ from the lowest 3 rotational states ($J \le 2$)
is found. For this cloud we derive an excitation temperature of $\simeq$ 70 K,
probably the kinetic temperature of the gas.
Another SMC component is visible at +160 km\,s$^{-1}$. 
Here we find unblended H$_2$\ absorption lines from levels $5 \le J \le 7$.
For this component we obtain an equivalent excitation
temperature of $>$ 2350 K and conclude that
this cloud must be highly excited by 
strong UV radiation from its energetic environment.

\keywords{Space Vehicles - ISM: molecules - Galaxies: ISM - 
          Magellanic Clouds: SMC - Stars: individual: HD\,5980
          - Ultraviolet: ISM}

\end{abstract}

\section{Introduction}

Investigations of interstellar molecular hydrogen only are possible in
the near IR in emssion and in the Far UV in absorption.
The {\it Copernicus} has been the first satellite which was able
to resolve UV absorption lines from galactic H$_2$\ in the range of
900 to 1200 \AA, and most of our knowledge about the physics of
interstellar H$_2$\ is based on studies with this instrument
(Spitzer et~al.\, 1974).
In contrast to observations in the near IR, FUV spectroscopy offers 
the possibility to investigate also the cool component of the gas 
in which the H$_2$\ molecule likely is the dominant constituent. 

More than twenty years after the first results of the {\it Copernicus}, 
the {\it ORFEUS\,}\ FUV telescope allows the measurement of H$_2$\ in all rotational 
states, not only in our own galaxy, 
but also in the interstellar matter of the two Magellanic Clouds.
This is of great importance because of the lower metal content 
and different gas to dust ratio in the Magellanic Clouds (see Koornneef 1984).
The {\it ORFEUS\,}\ 1m-telescope was launched for its second mission in Nov./Dec. 1996
with the {\it ASTRO-SPAS} space shuttle platform (Kr\"amer et~al.\, 1990).
Its two alternately operating spectrographs work in different wavelength
ranges with different resolutions.
The UCB spectrograph has a spectral resolution of 
$\lambda/\Delta\lambda \sim 5 \cdot 10^3$
in a wavelength range from 300 to 1200 \AA\ (Hurwitz et~al.\, 1998).
The echelle spectrograph operates between 912 and 1410 \AA\ with
a spectral resolution of $\lambda/\Delta\lambda \leq 10^4$.
A detailed description of this instrument and its performance is given by 
Kr\"amer et~al.\, (1990) and Barnstedt et~al.\, (1998). 

\begin{figure*}
\resizebox{\hsize}{!}{\includegraphics{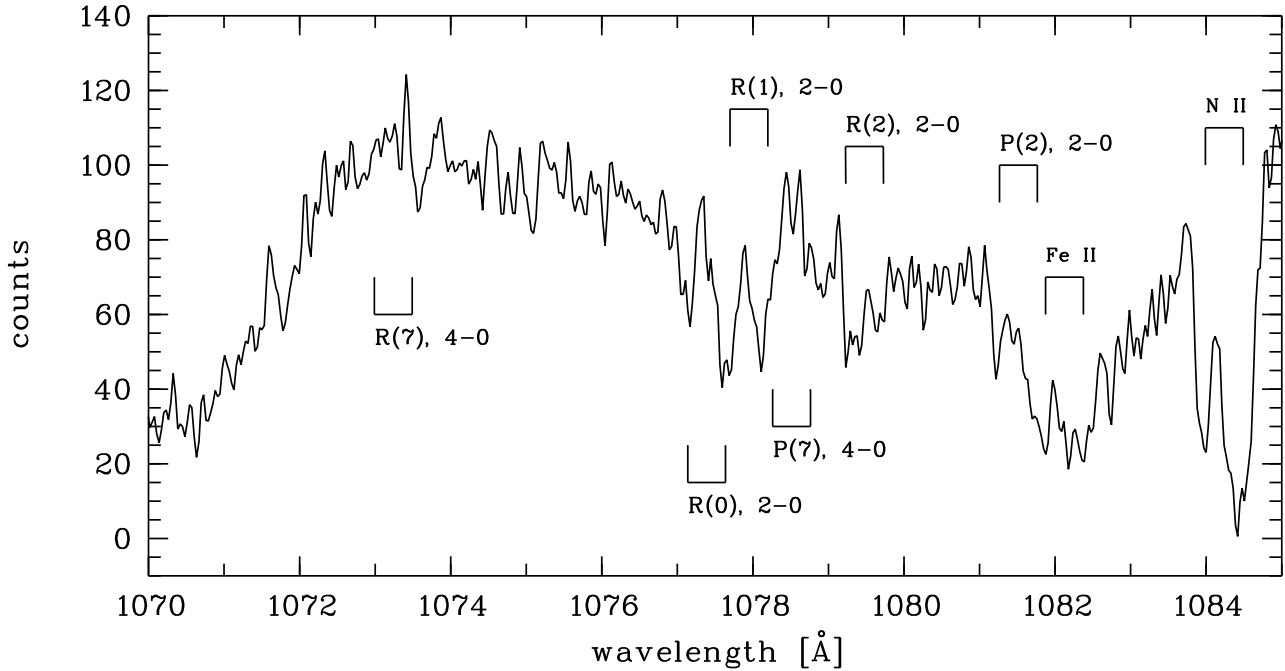}}
\caption[]{
The detected H$_2$\ lines in the displayed wavelength range
of the {\it ORFEUS\,}\ spectrum of HD\,5980 are identified.
The spectrum covers H$_2$\ absorption from the ground states as well from
rotational states as high as $J=7$. 
The vertical marks indicate the expected wavelength positions
for zero radial velocity
(galactic gas) and for 150 km\,s$^{-1}$\ (SMC gas). 
The spectral resolution is $\simeq 30$ km\,s$^{-1}$
}
\end{figure*}

Only one SMC target has been observed with the echelle during 
the {\it ORFEUS II} mission of Nov./Dec. 1996 : HD\,5980.
It is the brightest stellar object in the SMC, 
variable, and known for further extraordinary properties, 
as shown by Moffat et~al.\, (1998).
With its brightness and high temperature, 
as well as due to its relatively low extinction of 
$E(B-V)=0.07$ (Fitzpatrick \& Savage 1983, hereafter FS83), 
HD\,5980 provides us with a large UV flux and therefore is a good 
background source for the investigation of the SMC
foregroud gas along this line of sight.
HD\,5980 was also observed by {\it HUT} (Schulte-Ladbeck et~al.\, 1995)
but with a spectral resolution insufficient to resolve
H$_2$ .
The {\it IUE} spectrum of HD\,5980 shows the main SMC 
absorption in radial velocity near +140 km\,s$^{-1}$\ 
and one additional component at +300 km\,s$^{-1}$\ (FS83).
The {\it ORFEUS} spectrum, 
obtained when $V \simeq 10.8$ mag, shows both these components,
in particular also in O\,{\sc vi} (Widmann et~al.\, 1998).
McGee \& Newton (1986) showed the existence of neutral hydrogen in
21-cm emission at +123 km\,s$^{-1}$\ and at +163 km\,s$^{-1}$ .
We therefore expected to see H$_2$\ belonging to the neutral SMC gas
near those velocities, although due to the low
extinction towards HD\,5980 
the H$_2$\ absorption might be too weak and below our detection limit.

As shown by its near-IR emission, H$_2$\ exists in the SMC 
(Koornneef \& Israel 1985) as well as
in the LMC (Israel \& Koornneef 1991a, 1991b).
The present paper, together with the detection of H$_2$\ absorption profiles 
in the LMC (de\,Boer et~al.\, 1998), presents the first studies
of interstellar H$_2$\ in the Magellanic Clouds via 
high resolution FUV spectroscopy.

\section{Observations and data reduction}

HD\,5980 has been observed on 1+2 Dec. with a total observing time of 4800\,s.
After the main data reduction (see Barnstedt et~al.\, 1998 for
details) the individual echelle orders have been
filtered by a wavelet algorithm (Fligge \& Solanki 1997).
Our spectrum shows a signal-to-noise ratio (S/N) of $\simeq 25$ 
for wavelengths $>$ 1000 \AA; 
in the range $<$ 1000 \AA\ the lower sensitivity leads to a much poorer S/N.
The range below 1000 \AA\
is difficult to analyse, because the density
of H$_2$\ absorption lines is very high in that range.
Nearly all H$_2$\ lines are blended by other
transitions and atomic lines and their components.
In the spectral range between 1000 and 1130 \AA\ we
found 17 reasonably clean H$_2$\ absorption lines
with absorption at SMC velocities. 
A portion of the spectrum of HD\,5980 is shown in
Fig. 1. 

Some of the identified
H$_2$\ lines are plotted in the velocity scale (LSR) in Fig.\,2.
We detect absorption features from the lowest 8 rotational states. 
Our detection limit is $W_{\lambda} \simeq 25$ m\AA. 
Inspecting Fig.\,2 it is immediately clear that the H$_2$\ lines of the
low $J$ levels have absorption near +120 km\,s$^{-1}$\ and those
of high $J$ levels near +160 km\,s$^{-1}$ .

\begin{table}[t]
\caption[]{H$_2$\ column densities toward HD 5980 }
\begin{tabular}{lllll}
\hline\noalign{\smallskip}
 & Rotation  & $\log N(J)^a$ & $b$-value & Number of \\
 & level $J$ & [cm$^{-2}$] & [km\,s$^{-1}$]   & lines used\\
\noalign{\smallskip}
\hline\noalign{\smallskip}
Cloud A & 0 & 16.57 & 6 & 2 \\
        & 1 & 15.90 & 6 & 4 \\
\vspace{0.1cm}
        & 2 & 14.60 & 6 & 2 \\
Cloud B & 5 & 15.00 & 9 & 1 \\
        & 6 & 14.39 & 9 & 2 \\
        & 7 & 14.75 & 9 & 4 \\
\noalign{\smallskip}
\hline
\end{tabular}

\noindent
$^a$ for uncertainties see the error bars in Fig. 4
\end{table}

\begin{figure}
\resizebox{\hsize}{!}{\includegraphics{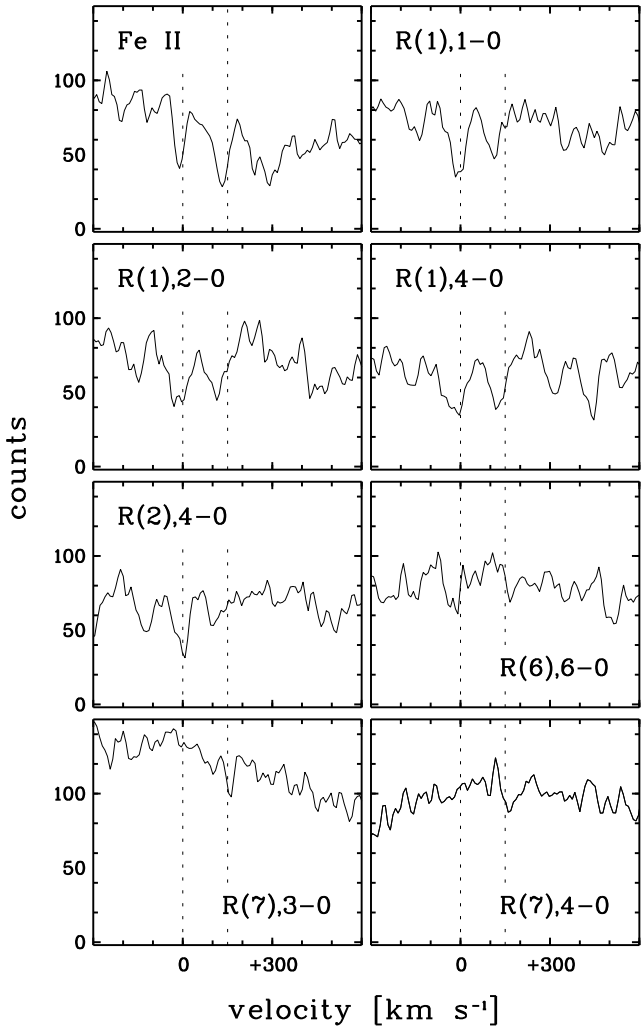}}
\caption[]{
A selection of H$_2$\ lines has been plotted in the velocity scale.
Galactic material absorbs at 0 km\,s$^{-1}$. 
The SMC H$_2$\ component for the lower roational states ($J \le3$)
is visible near +120 km\,s$^{-1}$, while the one for the higher states 
($5 \le J \le 7$) is present at +160 km\,s$^{-1}$ . 
The difference in the velocities indicates the presence of
{\it two} SMC clouds in our line of sight. At the top the profile of
Fe {\sc ii} at 1122.97 \AA\ is presented showing absorption near +140 km\,s$^{-1}$ .
The dotted lines indicate the velocities at 0 km\,s$^{-1}$\ and 150 km\,s$^{-1}$\ (LSR)
}
\end{figure}

\section{H$_2$\ column densities for the SMC gas}

The fact that we find H$_2$\ absorption in the SMC at $\simeq$ +120 km\,s$^{-1}$\ and
at $\simeq$ +160 km\,s$^{-1}$ , while the atomic component is visible near +140 km\,s$^{-1}$\,
is an indicator for the complexity of the SMC gas along this line of sight.
A detailed analysis of the atomic lines in {\it IUE} spectra is given
by FS83.
For the further analysis of the H$_2$\ absorption we call
the SMC component at +120 km\,s$^{-1}$\ {\it Cloud A} and the
one at +160 km\,s$^{-1}$\ {\it Cloud B}.

For $J=0,1,2$ only {\it Cloud A} shows strong H$_2$\ lines; 
a contribution by {\it Cloud B} might be present very weakly, 
but its absorption lies in the wings of the {\it Cloud A} and 
therefore is not visible.
For $J=3$ and $4$ the {\it Cloud B} absorption becomes stronger 
and overlaps with the {\it Cloud A} component, 
so that with our spectral resolution of $\simeq$ 30 km\,s$^{-1}$\ the H$_2$\ lines 
from both components combine to one single wider absorption feature.
At this stage of data analysis we do not attempt 
to separate the two clouds from each other 
in these intermediate $J$ level lines.

To derive H$_2$\ column densities for the SMC gas we have to construct curves of
growth for each cloud individually.
For that we use the $f$-values from Morton \& Dinerstein (1976).

\begin{figure}
\resizebox{\hsize}{!}{\includegraphics{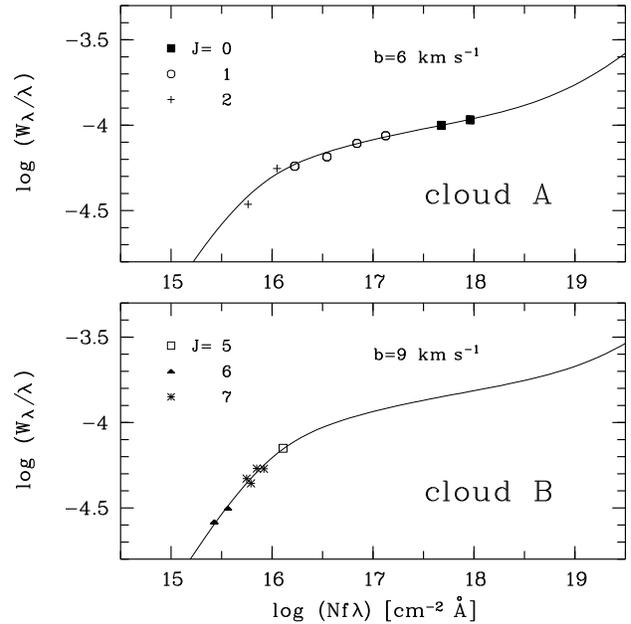}}
\caption[]{
The empirical curves of gowth for two SMC clouds are shown. 
{\it Cloud A} is at $\simeq$ +120 km\,s$^{-1}$\ and has no 
absorption from the higher rotational states, {\it Cloud B} is at 
$\simeq$ +160 km\,s$^{-1}$\ and has only marginal absorption from the
lower $J$ states
}
\end{figure}

For {\it Cloud A} we fitted 8 H$_2$\ lines of $J=0,1,2$ to
a curve of growth with a velocity dispersion of $b=5$ km\,s$^{-1}$ .
The total column density for $J \le 2$ 
is $N_{A}$(H$_{2})=4.6 \times 10^{16} $ cm$^{-2}$.

In {\it Cloud B} at +160 km\,s$^{-1}$\ we find H$_2$\ absorption for the 
higher states $J=5,6,7$ only. 
The existence of absorption from such high states 
indicates that the gas in {\it Cloud B} is highly excited.
The best fit for levels $J=5,6,7$ is given by a curve of growth
with $b=9$ km\,s$^{-1}$.

Both fits to the curve of growth are shown in Fig. 3, 
the obtained column densities can be found in Table 1.

For {\it Cloud B} we estimate the total amount of H$_2$\
by extrapolating the observed amount in the excited levels
based on the equivalent excitation temperature (see Section 4).
We so find $N_{B}$(H$_{2})=4.8 \times 10^{15} $ cm$^{-2}$.
The 21-cm emission of the neutral hydrogen shows a strong component
at +160 km\,s$^{-1}$\ (McGee \& Newton 1986). 

\section{H$_2$\ excitation state} 

Collisional excitation as well as the UV pumping are the processes responsible
for the population of the molecules excited states (Spitzer \& Zweibel 1974).
A measure for the kinetic state of the the gas is the equivalent excitation 
temperature,
which can be obained by fitting the given population density by a
Boltzmann distribution, as shown in Fig. 4.
The column densities $N(J)$ for both SMC clouds, devided by their statistical
weigth $g_J$, are plotted against the excitation energy $E_J$.

For {\it Cloud A} we derive an equivalent excitation temperature of $\simeq$
70 K. That means that the 3 lowest roational states of {\it Cloud A}
are most likely collisionally excited, so
that the derived temperature reflects the kinetic state of that
gas.

The situation looks different for {\it Cloud B}, where the fit for
$5 \le J \le 7$ leads to an equivalent excitation temperature of \,
$\simeq$ 2350 K. This temperature is not the kinetic
temperature of the gas, but gives the population of the excitation
states due to very strong UV pumping, likely caused by the high UV photon flux
environment of the SMC gas at +160 km\,s$^{-1}$ .

\begin{figure}
\resizebox{\hsize}{!}{\includegraphics{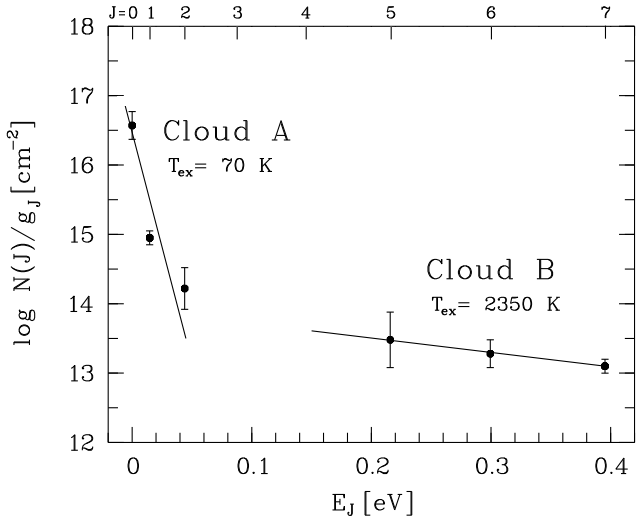}}
\caption[]{
The H$_2$\ column densities, devided by their statistical weight, are
plotted against the excitation energy. For {\it Cloud A} we 
obtain an equivalent excitation temperature of $\simeq$ 70 K for $J \le 2$, for 
{\it Cloud B} the states $5 \le J \le 7$ fit to a temperature
of $\simeq$ 2350 K. The derived temperatures are indicated 
by the solid lines. The errors shown are based on the 
uncertainties in the fits
of the curves of growth
}
\end{figure}

\section{Interpretation}

\subsection{{\it Cloud A}}

The excitation temperature for {\it Cloud A} is similar to values found 
in many studies of H$_2$\ gas in the Milky Way (Spitzer et al. 1974).
If we assume a galactic foreground reddening of $E(B-V)=0.02$ 
(McNamara \& Feltz 1980), the reddening in the SMC gas is $E(B-V)=0.05$
for this line of sight. 
The fact that we find absorption by H$_2$\ in the SMC gas toward a
star with that low extinction underlines that the gas-to-dust ratio in the
interstellar gas in the SMC is significantly higher than 
the Galactic value (FS83).
Neutral hydrogen has been detected in 21-cm emission towards HD\,5980 
at +123 km\,s$^{-1}$\ with a column density 
$N($H\,{\sc i}) $\simeq 1.39 \times 10^{21}$ cm$^{-2}$ 
(McGee \& Newton 1986).
Since this velocity is the same as that of {\it Cloud A} and 
since {\it Cloud A} shows little UV pumping, 
it lies likely well in front of HD 5980. 
We thus can compare $N$(H$_2$) andb $N$(H\,{\sc i}) and derive 
$f=2N($H$_2$$)/[N($H\,{\sc i})$ + 2N($H$_2$$)]= 6.6 \times 10^{-5}$
as the fraction of hydrogen nuclei in molecular form
in the interstellar gas of this cloud.
In view of the column density of $N$(H\,{\sc i}) at this velocity we
place this cloud in the foreground of the SMC.

\subsection{{\it Cloud B}}

The equivalent excitation temperature of $\simeq$ 2350 K
for {\it Cloud B} is the highest one ever seen in 
studies of H$_2$\ absorption and indicates a strong UV radiation
field in gas in the direct environment of {\it Cloud B},
probably from the target HD\,5980 itself. 
However, given the high excitation state and the large $N$(H\,{\sc i}) 
from 21 cm at the same velocity, 
we suggest that the main H\,{\sc i} emission comes from
gas {\it behind} HD\,5980 (see also FS83). 

HD 5980 is the visually brightest hot luminous stellar member of
the association NGC\,346, exciting the large H\,{\sc ii}\ region N\,66.
The star was in the late 1980s of WNE+OB type
with anomalously bright emission lines (Conti et~al.~1989).
HD 5980 is an eclipsing binary with a period of 19.3 days,
composed of a WN star and an O star with interacting stellar winds.
The star brightened slowly as of that time and the spectrum changed 
since then (see Moffat et~al.\, 1998 for the details).
During the ORFEUS measurements the star was of spectral type WN\,7. 

As shown in the {\it IUE} spectrum of HD\,5980, absorption by C\,{\sc iv} and 
Si\,{\sc iv} occurs at velocities near +150 km\,s$^{-1}$. 
FS83 conclude that
the bulk of this highly ionized gas is formed by stellar 
photoionization of HD\,5980.
Since we detect H$_2$\ near +160 km\,s$^{-1}$\ in {\it Cloud B} 
it is reasonable to believe that the molecular gas belongs
to the component found by FS83.
The H$_2$\ of {\it Cloud B} 
is highly influenced by the large UV flux of HD\,5980
and of NGC\,346 as a whole, and thus likely resides
in the immediate surrounding of the star and the cluster.

\section{Concluding remarks}

The {\it ORFEUS} FUV spectrum of HD\,5980 shows absorption
by interstellar H$_2$\ in the SMC.
Two velocity components have been detected belonging to the
SMC gas. H$_2$\ gas at +120 km\,s$^{-1}$\ is predominantly collisionally excited, 
as indicated by the lowest three rotational states. 
It is similar to galactic H$_2$ gas. 
H$_2$\ gas seen at +160 km\,s$^{-1}$\ is highly
excited, probably due to UV pumping by the abundant UV photons
from HD\,5980 and NGC\,346.
These findings represent, 
together with the {\it ORFEUS\,} detection of H$_2$\ in the
LMC (de\,Boer et~al.\, 1998), the first
studies of H$_2$\ in absorption in the
Magellanic Clouds.

\begin{acknowledgements}
ORFEUS could only be realized with the support of all our German and
American colleagues and collaborators.
The ORFEUS project was supported by DARA grant WE3 OS 8501, WE2 QV 9304 and
NASA grant NAG5-696. PR is supported by DARA grant 50\,QV\,9701\,3
\end{acknowledgements}

{}
\end{document}